\documentclass[journal]{IEEEtran}
\usepackage[dvips]{graphicx}
\usepackage{subfig}
\usepackage{amsfonts,amssymb,amsmath,amsbsy}
\usepackage{color}

\def\beq{\begin{equation}}
\def\eeq{\end{equation}}
\def\bit{\begin{itemize}}
\def\eit{\end{itemize}}
\def\beqq{\begin{eqnarray}}
\def\eeqq{\end{eqnarray}}

\def\ie{\textit{i.e.}}

\def\E{\mathbb{E}}

\def\Y{\mathbf{Y}}
\def\y{\mathbf{y}}
\def\A{\mathbf{A}}

\def\bP{\mathbf{P}}

\DeclareFontFamily{U}{euc}{}
\DeclareFontShape{U}{euc}{m}{n}{<-6>eurm5<6-8>eurm7<8->eurm10}{}%
\DeclareSymbolFont{AMSc}{U}{euc}{m}{n} 
\DeclareMathSymbol{\umu}{\mathord}{AMSc}{"16}
\def\p{\mathbf{p}}

\def\brho{\boldsymbol\rho}
\def\bbeta{\boldsymbol\beta}
\newcommand{\ind}[1]{\mathbf{1}_{\{#1\}}}
\def\numax{\nu_u}
\def\numin{\nu_l}
\def\y{\mathbf{y}}
\def\Y{\mathbb{Y}}
\def\A{\mathbb{A}}

\newtheorem{lem}{Lemma}

\usepackage{ifthen}

\newcounter{sgSingleColumn}
\ifthenelse{\value{sgSingleColumn}=1}{}{\usepackage[nomarkers,nolists]{endfloat}}

\renewcommand{\IEEEQED}{}

\usepackage{algorithm2e}

\allowdisplaybreaks

\begin{document}
\title{Cognitive Coexistence between \\Infrastructure and Ad-hoc Systems}
\author{Stefan~Geirhofer,~\IEEEmembership{Student Member,~IEEE,}
        Lang~Tong,~\IEEEmembership{Fellow,~IEEE,}
        and~Brian~M.~Sadler,~\IEEEmembership{Fellow,~IEEE}
\thanks{Manuscript received November 21, 2008.  This work has been submitted to the IEEE for possible publication.  Copyright may be transferred without notice, after which this version may no longer be accessible.}
\thanks{This work is supported in part by the U.S.~Army Research Laboratory under the Collaborative Agreement DAAD19-01-2-0011.  The U.S.~Government is authorized to reproduce and distribute reprints for Government purposes notwithstanding any copyright notation thereon.  This work has been presented in part at the IEEE Military Communications Conference, San Diego, CA, November 2007.}
\thanks{Stefan Geirhofer and Lang Tong are with the School of Electrical and Computer Engineering, Cornell University, Ithaca, NY, 14583 (e-mail: \{sg355, lt35\}@cornell.edu).  Brian~M.~Sadler is with the U.S.~Army Research Laboratory, Adelphi, MD, 20783 (e-mail: bsadler@arl.army.mil).}}

\markboth{Submitted to IEEE Transactions on Wireless Communications}%
{Submitted to IEEE Transactions on Wireless Communications}

\maketitle

\IEEEpeerreviewmaketitle

\def\gambar{\bar{\gamma}}
\def\nubar{\bar{\nu}}
\def\eg{\textit{e.g.}}
\def\vrho{\brho}
\begin{abstract}
The rapid proliferation of wireless systems makes interference management more and more important.  This paper presents a novel cognitive coexistence framework, which enables an infrastructure system to reduce interference to ad-hoc or peer-to-peer communication links in close proximity.  Motivated by the superior resources of the infrastructure system, we study how its centralized resource allocation can accommodate the ad-hoc links based on sensing and predicting their interference patterns.

Based on an ON/OFF continuous-time Markov chain model, the optimal allocation of power and transmission time is formulated as a convex optimization problem and closed-form solutions are derived. The optimal scheduling is extended to the case where the infrastructure channel is random and rate constraints need only be met in the long-term average.  Finally, the multi-terminal case is addressed and the problem of optimal sub-channel allocation discussed.  Numerical performance analysis illustrates that utilizing the superior flexibility of the infrastructure links can effectively mitigate interference.
\end{abstract}


\begin{figure*}[!t]
\centering
\subfloat[System setup]{\includegraphics{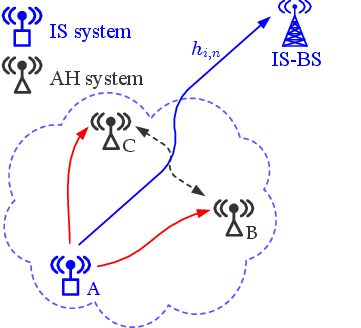}
\label{fig:1a-SystemSetupA}}
\subfloat[Time/frequency setup]{\includegraphics{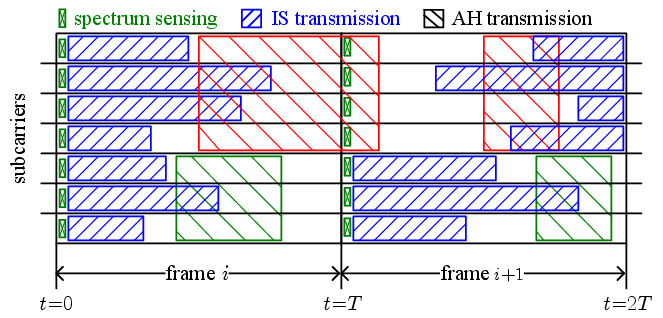}
\label{fig:1b-SystemSetupB}}
\caption{System setup.  An infrastructure link allocates power and transmission time such as to minimize interference to close-by ad-hoc networks.  The interference-aware resource allocation is based on detecting and predicting the ad-hoc system's temporal activity. }
\label{fig:1-SystemSetup}
\end{figure*}

\section{Introduction}
\label{sec:1--Introduction}
The rapid growth of wireless networks makes interference an important performance impediment and motivates a careful study of coexistence.  In unlicensed bands, where there is a lack of coordination among heterogeneous technologies, traditional forms of coexistence are based on statically separating systems, for example by having them operate in disjoint frequency bands.  However, such static approaches are inefficient and not able to accommodate the projected growth in deployments.

The high interference conditions which are prevalent in unlicensed bands are in contrast to a fairly low average utilization, which results from the burstiness of traffic that is supported by these systems and the random medium access behavior typically employed.  This contrast motivates dynamic approaches to interference management which can leverage unused spectrum opportunities.

Cognitive radio presents a new framework for analyzing this problem.  Reconfigurable radio platforms enable a dynamic adaptation of transmission parameters based on monitoring the radio environment.  In this way it is possible to reduce interference by detecting and predicting temporal activity patterns and avoiding transmission overlaps.  Ultimately this may lead to a new \emph{cognitive coexistence} paradigm.

This paper addresses a special case of cognitive coexistence involving two different types of networks: an infrastructure (IS) wide area network that shares spectrum with local, ad-hoc (AH) or peer-to-peer systems.  Motivated by the superior communication resources of the IS system, we analyze how its flexible, centralized resource allocation can accommodate the AH links based on sensing and predicting their interference patterns.  Despite adapting its resource allocation based on sensing results, the IS system minimizes interference to the AH network subject to maintaining a specified quality-of-service (QoS) level for its users.

This approach is different from typical dynamic spectrum access (DSA) problem formulations in which a secondary system exploits spectrum opportunities left over by a primary system, subject to the constraint that no significant interference is created (see \cite{Zhao&Tong:06SPM} for a review).  While both approaches are hierarchical, spectrum property rights in DSA mandate that secondary users adapt to spectrum licensees.  In contrast, in cognitive coexistence it is possible to use the flexibility of the primary system to accommodate a secondary (perhaps lower priority) system while maintaining a desired primary system performance level. This could be viewed as a ``best-effort'' approach toward interference management and coexistence.

This framework is relevant to a number of practical setups.  For example, the convergence of wide and local area networks has received increasing interest.  The coexistence of IEEE~802.16 and IEEE~802.11 systems in unlicensed bands is such an example of practical importance \cite{Berlemann&Hoymann&Hiertz&Walke:06:PIMRC,Fu&Ma&Zhang:07:WCNC}.  Similarly, incorporating peer-to-peer connectivity into cellular networks \cite{Fitzek&Katz07Book} or accommodating femto-cell base stations \cite{Claussen&Ho&Samuel:08:BellLabsTechJ} are potential scenarios where this work could be applied.  Yet other applications arise in the military domain, where the coexistence of high and low priority links is a fundamental concern \cite{Burbank&Kasch:00:MILCOM}.

\subsection{Main contribution}
\label{sec:1a-MainContribution}
This paper addresses the coexistence of two wireless networks that operate on different spatial scales: a longer range IS network that interferes with local AH links.  We address the question of whether the IS network can accommodate surrounding AH links by allocating power and transmission time judiciously.  Specifically, we make the following contributions:
\begin{itemize}
\item The problem of optimal power and transmission time allocation is formulated as a convex program and the optimal frame-level solution is derived.  Based on optimality conditions, a solution algorithm with guaranteed convergence and low complexity is introduced.  The structure of the optimal solution is studied to provide further intuition.
\item The frame-level problem is relaxed to the case of average rate constraints, in which statistical knowledge of the activity patterns of the AH users and the IS channel coefficients are used to allocate resources in both frequency and time.
\item Finally, a scenario with multiple IS users is considered in which the IS base station allocates sub-channels based on average interference metrics and users perform optimal power and transmission time allocation based on sensing results.  A comparison with conventional sub-channel allocation methods shows that heuristics may yield performance close to optimal.
\end{itemize}
These contributions are corroborated by numerical results which demonstrate that judicious transmission time allocation can mitigate interference effectively.

\subsection{Related work}
\label{sec:1b-RelatedWork}
Optimal resource management in multiuser multicarrier wireless systems has been well-studied for both downlink and uplink cases; see \cite{Wong&Evans:08,Kim&Han&Kim:05:COML,Wong&Cheng&Letaief&Murch:99:JSAC} for an overview of the topic.  In cognitive radio networks, optimal resource allocation is more challenging because it needs to incorporate interference constraints, which protect the primary system from harmful interference.  Typical formulations aim at finding a power and sub-channel allocation which maximizes the throughput of the cognitive radio system while meeting interference and power constraints.  Recent work in this area includes \cite{Peng&Wang&Lu&Wang:07:WiCom,Wang&Zhao&Xiao&Etal:07:GlobeCom}.  In addition to meeting interference constraints, spectrum sharing and self-coexistence within the cognitive radio network also need to be addressed.  Contributions in this area include \cite{AcharyaYates:07:ICC,Hoang&Liang:06:PIMRC,Wang&Liu:05VTC,Zheng&Peng05ICC}.
Within the framework of cognitive coexistence, the optimal power allocation based on knowledge of the interference channel has been addressed in \cite{Geirhofer&Tong&Sadler:08:Globecom} by the authors of this paper.  Previous work on improving the coexistence among local and personal area networks includes \cite{Geirhofer&Tong&Sadler:07ComMag, Geirhofer&Tong&Sadler:08JSAC} in which a cognitive frequency hopping protocol is derived based on temporal activity models.  To the best of the authors' knowledge, interference-aware power and transmission time allocation  based on predicting temporal activity patterns has not been addressed before.

\subsection{Organization and notation}
\label{sec:1c-Organization}
The rest of this paper is organized as follows.  After introducing the problem setup in Sec.~\ref{sec:2--ProblemFormulation} the optimal frame-level allocation is derived in Sec.~\ref{sec:3--FrameLevelAllocation}.  The results are extended to the average rate case in Sec.~\ref{sec:4--AverageResourceAllocation} and sub-channel assignment is considered in Sec.~\ref{sec:5--MultiTerminalAllocation}.  Throughout this paper notation is fairly standard.  Vectors are typeset in boldface.  For an event $\mathcal{X}$, the indicator function $\mathbf{1}_{[\mathcal{X}]}$ is equal to one if $\mathcal{X}$ occurs and zero otherwise.  The notation $(\cdot)^+$ is used to abbreviate $\max\{0,\cdot\}$.

\section{Problem Formulation}
\label{sec:2--ProblemFormulation}

\subsection{System setup}
\label{sec:2a-Organization}
The system setup is shown in Fig.~\ref{fig:1-SystemSetup}.  We consider an IS system, which consists of a base station and a single client (the multi-terminal case will be addressed in Sec.~\ref{sec:5--MultiTerminalAllocation}). The uplink transmissions of this client may strongly interfere with local transmissions of one or multiple AH networks surrounding the client.  For this setup, the problem of optimally assigning power and transmission time at the IS client, such as to minimize interference to the AH links is analyzed.  The time/frequency behavior of both systems is shown in Fig.~\ref{fig:1b-SystemSetupB}.

\subsubsection*{Ad-hoc network}
The AH network consists of a set of AH nodes which operate in a frequency band that overlaps with the IS system.  As depicted in Fig.~\ref{fig:1b-SystemSetupB} there can be multiple AH networks which operate in non-overlapping bands that each overlap with a certain set of IS sub-channels.  It is assumed that the partitioning of the AH bands is fixed and that the temporal activity of different bands is statistically independent.

We model the time behavior of each AH band by a two-state ON/OFF continuous time Markov chain (CTMC).  The holding times in both ON and OFF state are exponentially distributed with parameters $\mu$ for the ON state and $\lambda$ for the OFF state.  Therefore, if an AH link is detected to be in a certain state at time $t_0$, then its transition matrix for time $t_0+\tau$ is given by
\beq
\label{eq:CTMCTransitionMatrix}
\bP(\tau) = \frac{1}{\lambda+\mu}
\left[
  \begin{array}{cc}
    \mu+\lambda e^{-(\lambda+\mu)\tau} & \lambda - \lambda e^{-(\lambda+\mu)\tau} \\
    \mu-\mu e^{-(\lambda+\mu)\tau} & \lambda + \mu e^{-(\lambda+\mu)\tau} \\
  \end{array}
\right],
\eeq
which follows directly from the definition of a CTMC \cite[p.391]{Resnick92}.  Therefore, the probability of an AH link being ON at time $t_0+\tau$, conditioned on having it observed in the ON (OFF) state at time $t_0$ is given by the lower right (upper right) entry in the matrix above.

Modeling AH links based on a two-state CTMC approximates the carrier sense random medium access typically employed in such systems.  This modeling approach has been used in related publications \cite{Geirhofer&Tong&Sadler:08JSAC,Zhao&Geirhofer&Tong&Sadler:07SPsub} and solidified by a measurement-based analysis of WLAN traffic \cite{Geirhofer&Tong&Sadler:07ComMag}.  The CTMC assumption strikes a good tradeoff between model accuracy and the analytical tractability that is needed in the subsequent sections.

\subsubsection*{Infrastructure system}
The IS system operates in the same frequency band as the AH network and evolves in frames of fixed duration $T$.  At the beginning of each frame, spectrum sensing is used to detect the ON/OFF activity of the AH bands, and based on the sensing outcome, power and transmission time are assigned; see Fig.~\ref{fig:1b-SystemSetupB}.

Throughout the paper we assume \emph{perfect sensing}, that is, sensing outcomes are always accurate and the overhead associated with sensing is negligible.\footnote{%
In contrast to many DSA setups, where very weak signals need to be detected, the proximity of AH terminals to the IS client leads to moderate to high average signal-to-noise-ratios.  This facilitates the sensing task and enables us to employ simple methods such as energy detection.}
Due to the proximity of IS and AH systems, the detection task is conceptually similar to the carrier sensing employed in systems such as IEEE~802.11.

Based on the sensing result at the beginning of each frame, the IS system allocates power and transmission time on a sub-channel basis.  This is conceptually similar to the allocation of time/frequency resource blocks in broadband cellular systems based on OFDMA.  The case where a subset or even all sub-channels need to share the same timing allocation (for example when transmissions in the entire band can only be turned on or off) has worse performance in general.  Nevertheless, a similar solution approach remains applicable.

The IS system minimizes interference subject to maintaining rate requirements for its client.  The rate that is supported by a specific sub-channel is modeled based on a channel capacity formulation,
\beq
\label{eq:AchievableRate}
\sum_n \rho_n\log\left(1+\kappa\frac{p_n|h_n|^2}{\rho_n N_0}\right)=
\sum_n \rho_n\log\left(1+\frac{p_n\beta_n}{\rho_n}\right),
\eeq
where $\p=[p_1,\ldots,p_N]^T$ denotes the power allocation, $\brho = [\rho_1,\ldots,\rho_N]^T$ represents the transmission time allocation, $N_0$ is the noise power, $\kappa$ a normalization factor, and $\beta_n$ is introduced for notational convenience%
\footnote{The above formulation encompasses a channel capacity formulation (for $\kappa=1$) as well as the case of variable-rate M-QAM in which case $\kappa=1.5/(-\ln\mathsf{BER})$ is chosen such that a target $\textsf{BER}$ is met \cite{Qiu&Chawla}.}.

\subsection{Interference metrics and scheduling assumptions}
\label{sec:2b-InterferenceMetrics}
The interference between IS and AH networks is modeled by the average temporal overlap between both systems.  Based on the sensing result at the beginning of the frame and knowledge of the CTMC parameters of the AH links, transmission time and power are allocated.

The allocation of transmission time consists of specifying duration and placement of the transmission within the current frame.  We first show that it is optimal to transmit at the beginning (the end) of the frame if the sensing outcome is idle (busy).

\begin{lem}
Assume that a $\rho$ fraction of transmission time needs to be allocated to a sub-channel, on which the AH user's ON/OFF behavior is modeled by the CTMC (\ref{eq:CTMCTransitionMatrix}).  Based on a sensing outcome at the beginning of the frame, the minimum expected overlap with the ON period of the AH user is achieved by
    \bit
    \item transmitting at the beginning of the frame (\ie, during $[0,\rho T]$) if the sensing outcome was idle and
    \item transmitting at the end of the frame (\ie, during $[(1-\rho)T,T]$ if the sensing outcome was busy.
    \eit
\end{lem}
\begin{proof}
see appendix.
\end{proof}

Based on Lemma~1, we derive the expected time overlap between IS and AH transmissions, conditioned on the sensing result $y\in\{0,1\}$ at the beginning of the frame.  Consider a sub-channel $n$, which overlaps with AH band $i=g(n)$.  Then, the activity of AH user $i$ is given by the CTMC $\{X_i(\xi), \xi\geq0\}$ with parameters $\lambda_i$ and $\mu_i$.  Transmitting for a $\rho$ fraction of the frame, leads to the expected time overlap
\ifthenelse{\value{sgSingleColumn}=1}{%
\begin{multline}
\label{eq:CostFunction-Idle}
\phi_{n,0}(\rho) = \frac{1}{T}\E\left\{\int\limits_0^{\rho T}\ind{X_i(\xi)=1}d\xi\bigg|X(0)=0\right\}\\ = \frac{1}{T}\int\limits_0^{\rho T}\Pr(X(\xi)=1|X(0)=0)d\xi
\end{multline}}{
\beq
\label{eq:CostFunction-Idle}
\phi_{n,0}(\rho) = \frac{1}{T}\E\left\{\int\limits_0^{\rho T}\ind{X_i(\xi)=1}d\xi\bigg|X(0)=0\right\} = \frac{1}{T}\int\limits_0^{\rho T}\Pr(X(\xi)=1|X(0)=0)d\xi
\eeq}
if the sensing result was idle.  By substituting (\ref{eq:CTMCTransitionMatrix}) it is then easy to show that
\beq
\phi_{n,0}(\rho) = \tfrac{\lambda_i}{(\lambda_i+\mu_i)T}\left(\rho T + \tfrac{1}{\lambda_i+\mu_i}(e^{-(\lambda_i+\mu_i)\rho T}-1)\right).
\eeq
In the case of a busy sensing result we obtain
\ifthenelse{\value{sgSingleColumn}=1}{%
\begin{multline}
\label{eq:CostFunction-Busy}
\phi_{n,1}(\rho) =\\ \tfrac{\lambda_i}{(\lambda_i+\mu_i)T}\left(\rho T +\tfrac{\mu_i/\lambda_i}{\lambda_i+\mu_i}e^{-(\lambda_i+\mu_i)T}(e^{(\lambda_i+\mu_i)\rho T}-1)\right).
\end{multline}}{
\beq
\label{eq:CostFunction-Busy}
\phi_{n,1}(\rho) =\\ \tfrac{\lambda_i}{(\lambda_i+\mu_i)T}\left(\rho T +\tfrac{\mu_i/\lambda_i}{\lambda_i+\mu_i}e^{-(\lambda_i+\mu_i)T}(e^{(\lambda_i+\mu_i)\rho T}-1)\right).
\eeq
}

This derivation made use of the fact that the sensing results of two IS sub-channels are either perfectly correlated (if they overlap with the same AH band) or statistically independent (if they overlap with different bands).  Therefore, the prediction performance of a specific sub-channel cannot be improved by using sensing results from other sub-channels.

\begin{lem}
The functions $\phi_{n,0}(\rho)$ and $\phi_{n,1}(\rho)$ are strictly convex and increasing in $\rho$.
\end{lem}
\begin{proof}
Both $\phi_{n,0}(\rho)$ and $\phi_{n,1}(\rho)$ are nonnegative linear combinations of a convex and a strictly convex function.  One is linear, the other an exponential function with nonzero exponent.  The monotonicity can easily be verified by differentiation.
\end{proof}

\section{Optimal Frame-Level Allocation}
\label{sec:3--FrameLevelAllocation}

Consider a single IS client, which minimizes the time overlap between IS and AH transmissions subject to maintaining a rate constraint across the IS channel to the IS base station.  Mathematically, this leads to problem \textbf{P1}
\begin{align}
\min\limits_{\p,\brho}\ & \sum_n \phi_{n,y_n}(\rho_n)\label{eq:Overlap-Objective} \\
\text{s.t.}\ & \sum_n \rho_n\log\left(1+\frac{p_n\beta_n}{\rho_n}\right) \geq R \label{eq:Overlap-RateConstr}\\
& \sum_n p_n  \leq P\label{eq:Overlap-PowerConstr}\\
& p_n\geq 0,\quad 1\leq n\leq N\label{eq:Overlap-RegularityP}\\
& 0\leq\rho_n\leq 1,\quad 1\leq n\leq N,\label{eq:Overlap-RegularityRho}
\end{align}
with rate constraint (\ref{eq:Overlap-RateConstr}) and power constraint (\ref{eq:Overlap-PowerConstr}).  It is straightforward to show that \textbf{P1} is a convex optimization problem since the objective function is convex (by Lemma~2), the rate constraint (once rewritten in standard form) is convex by the perspective property \cite{Boyd&Vandenberghe}, and all other constraints are linear.

A solution to \textbf{P1} can be found by general solution techniques in polynomial time \cite{Boyd&Vandenberghe}.  For this specific problem, however, it is possible to show a special structure that enables us to gain further insight into the problem.

\subsection{Optimality conditions and solution structure}
\label{sec:3a-OptimalityConditionsAndSolutionStructure}
The solution structure is obtained by introducing Lagrange multipliers $\gamma$ and $\epsilon$ for the rate and power constraint, respectively.  This leads to the Lagrangian
\ifthenelse{\value{sgSingleColumn}=1}{%
\begin{multline}
\label{eq:Lagrangian}
L(\p,\brho; \gamma,\epsilon) = \sum_n\phi_{n,y_n}(\rho_n)+\\
\gamma\left[R-\sum_n\rho_n\log\left(1+\frac{p_n\beta_n}{\rho_n}\right)\right]
+\epsilon\left[\sum_n p_n-P\right].
\end{multline}}{
\beq
\label{eq:Lagrangian}
L(\p,\brho; \gamma,\epsilon) = \sum_n\phi_{n,x_n}(\rho_n)+
\gamma\left[R-\sum_n\rho_n\log\left(1+\frac{p_n\beta_n}{\rho_n}\right)\right]
+\epsilon\left[\sum_n p_n-P\right].
\eeq}
The Karush-Kuhn-Tucker (KKT) optimality conditions are then given by the constraints (\ref{eq:Overlap-RateConstr})-(\ref{eq:Overlap-RegularityRho}) of \textbf{P1}, nonnegativity constraints for the Lagrange multipliers, $\gamma\geq0$, $\epsilon\geq0$, the slackness conditions
\beqq
\gamma\left[R-\sum_n\rho_n^*\log\left(1+\frac{p_n^*\beta_n}{\rho_n^*}\right)\right] & = & 0 \label{eq:Slackness-R}\\
\epsilon\left[\sum_n p_n^* -P\right] & = & 0,\label{eq:Slackness-P}
\eeqq
the condition
\beq
\label{eq:OptimalityCondition-P}
\frac{\partial L(\p,\brho;\gamma,\epsilon)}{\partial p_n}\bigg|_{p_n=p^\ast_n}
\left\{
\begin{array}{ll}
  =0, & p_n^*>0 \\
  >0, & p_n^*=0
\end{array}
\right.,
\eeq
and
\beq
\label{eq:OptimalityCondition-Rho}
\frac{\partial L(\p,\brho;\gamma,\epsilon)}{\partial \rho_n}\bigg|_{\rho_n=\rho^\ast_n}
\left\{
\begin{array}{ll}
  >0, & \rho_n^\ast=0\ \\
  =0, & \rho_n^\ast\in(0,1) \\
  <0, & \rho_n^\ast=1
\end{array}
\right..
\eeq
Expressions (\ref{eq:OptimalityCondition-P}) and (\ref{eq:OptimalityCondition-Rho}) can be understood on an intuitive level by noting that for $p_n^\ast$ and $\rho_n^\ast$ to be minimizers of $L(\p,\brho;\gamma,\epsilon)$, their partial derivative must equal zero unless they lie on the boundary of the feasible set.

\subsubsection*{Solution structure for $p_n^*$}
By substituting (\ref{eq:Lagrangian}) into (\ref{eq:OptimalityCondition-P}) and solving for $p_n^\ast$ we arrive at
\beq
\label{eq:OptSolutionStructure-P}
p_n^\ast = \rho_n\left(\nu-\frac{1}{\beta_n}\right)^+,
\eeq
where $\nu:=\gamma/\epsilon$ has been introduced to simplify notation in what follows.  For any fixed value of $\rho_n$, (\ref{eq:OptimalityCondition-P}) represents a water filling solution \cite{Cover&Thomas}.

\subsubsection*{Solution structure for $\rho_n^*$}
The optimal transmission time allocation is obtained by substituting (\ref{eq:Lagrangian}) and (\ref{eq:OptSolutionStructure-P}) into (\ref{eq:OptimalityCondition-Rho}).  For an idle sensing result, $y_n=0$ we obtain,
\beq
\label{eq:OptSolutionStructure-Rho}
\rho_n^\ast = \left\{
\begin{array}{ll}
  \frac{1}{(\lambda_i+\mu_i)T}\log\frac{1}{1-\frac{\lambda_i+\mu_i}{\lambda_i}\gamma h_n(\nu)}, & \gamma h_n(\nu)\leq \zeta_{0,i}\\
  1, & \text{o.w.}
\end{array}
\right.,
\eeq
where $\zeta_{0,i}=\lambda_i/(\lambda_i+\mu_i)(1-\exp(-(\lambda_i+\mu_i)T)$ and $i=g(n)$ denotes the AH sub-band that overlaps with sub-channel $n$.  In the above equation we have defined
\beq
h_n(\nu) := [\log(\nu\beta_n)]^+ - \frac{(\nu\beta_n-1)^+}{1+(\nu\beta_n-1)^+}
\eeq
to simplify notation.  Similarly, we can obtain the solution structure for the case of a busy sensing result, $y_n=1$,
\beq
\label{eq:OptSolutionStructure-RhoBusy}
\rho_n^* = \left\{
\begin{array}{ll}
  0, & \gamma h_n(\nu)<\zeta_{1,i} \\
  1+\frac{\log\left(\frac{\lambda_i+\mu_i}{\mu_i}\gamma h_n(\nu)-\frac{\lambda_i}{\mu_i}\right)}{(\lambda_i+\mu_i)T}, & \zeta_{1,i}\leq \gamma h_n(\nu)\leq 1 \\
  1, & \gamma h_n(\nu)>1,
\end{array}
\right.,
\eeq
where $\zeta_{1,i}=\lambda_i/(\lambda_i+\mu_i)(1+\mu_i/\lambda_i\exp(-(\lambda_i+\mu_i)T)$. Note that these closed-form expressions depend on the Lagrange multipliers only through the term $\gamma h_n(\nu)$ which does not depend on the AH activity parameters $\lambda_i$ and $\mu_i$.  Further, the transmission time allocations \eqref{eq:OptSolutionStructure-Rho}-\eqref{eq:OptSolutionStructure-RhoBusy} are monotonic with respect to this term.

\subsection{Iterative solution algorithm for $\gamma$ and $\nu$}
\label{sec:3b-IterativeSolutionAlgorithm}

To find the optimal power and transmission time allocation based on the above closed-form expressions, we present an algorithm for finding the pair $[\gamma^*,\nu^*]$, which corresponds to the optimal solution of \textbf{P1}.

For any pair $[\gamma,\nu]$ the power allocation $\p(\gamma,\nu)$ and transmission time allocation $\brho(\gamma,\nu)$ define the optimal solution to \textbf{P1} with modified rate constraint
\beq
\label{eq:GammaProof_R}
R(\gamma,\nu):=\sum_n\rho_n(\gamma,\nu)[\log(\nu\beta_n)]^+
\eeq
and modified power constraint
\beq
\label{eq:GammaProof_P}
P(\gamma,\nu):=\sum_n\rho_n(\gamma,\nu)\left(\nu-\frac{1}{\beta_n}\right)^+,
\eeq
where $\rho_n(\gamma,\nu)$ is given by (\ref{eq:OptSolutionStructure-Rho}) or (\ref{eq:OptSolutionStructure-RhoBusy}) (depending on the sensing result).  The fact that this solution is optimal for rate constraint $R(\gamma,\nu)$ and power constraint $P(\gamma,\nu)$ follows directly from the KKT optimality conditions, which are necessary and sufficient for convex optimization problems \cite{Boyd&Vandenberghe}.

Based on the above, finding the pair $[\gamma^*,\nu^*]$ corresponding to the given rate constraint $R$ and power constraint $P$ could theoretically be performed by searching all pairs $[\gamma,\nu]$.  In this section we show, however, that $R(\gamma,\nu)$ and $P(\gamma,\nu)$ exhibit some monotonicity which enables us to use the bisection method for finding $[\gamma^*,\nu^*]$ with guaranteed convergence and low complexity.  We first study the case of keeping $\gamma$ fixed and adjusting $\nu$ such that the rate constraint is met with equality.  Then, we show that the allocated sum power decreases with $\gamma$.

\subsubsection*{Adjusting $\nu$ to meet the rate constraint}
We first consider the case of adjusting $\nu$ such that $R(\gamma,\nu)=R$ while keeping $\gamma$ fixed.  It is easy to verify that for any $n$, $h_n(\nu)$ is nondecreasing in $\nu$.  Therefore, $\rho_n(\gamma,\nu)$ increases with $\gamma$ as well as can be seen from (\ref{eq:OptSolutionStructure-Rho}) and (\ref{eq:OptSolutionStructure-RhoBusy}).  Further, since $\rho_n(\gamma,\nu)$ increases with $\nu$ for fixed $\gamma$, so does $R(\gamma,\nu)$.  We can exploit this property to find the $\nu$ for which $R(\gamma,\nu)=R$ by the bisection method. First, we can find upper and lower bounds, $\nu_u$ and $\nu_l$, for this value. These bounds are guaranteed to exist since $R(\gamma,\nu)\to\infty$ for $\nu\to\infty$ and $R(\gamma,\nu)\to0$ for $\nu\to0$.  Once these bounds have been obtained the bisection method iteratively finds $\nu^*$ with guaranteed convergence.

\subsubsection*{Adjusting $\gamma$ to meet the power constraint}
Having obtained an algorithm for finding $\nu$ for arbitrary $\gamma$ such that the rate constraint is satisfied, we study the behavior of the power constraint as $\gamma$ is adjusted.  As $\gamma$ is varied, we continue to adjust $\nu$ such that the rate constraint is satisfied at all times.  The pair of Lagrange multipliers is therefore given by $[\gamma,\nu^*(\gamma)]$.

The slackness conditions imply that at the optimal solution both rate and power constraints are met with equality.  From (\ref{eq:GammaProof_R}) we observe that decreasing $\gamma$ requires increasing $\nu$ in order to continue meeting the rate constraint. %
Further, decreasing $\gamma$ reduces the objective function because $\log(\hat{\nu}\beta_n)\geq\log(\nu\beta_n)$ for $\hat{\nu}\geq\nu$, enables us to reduce $\rho_n(\gamma,\nu^*(\gamma))$ for at least some $n$.

Since decreasing $\gamma$ requires increasing $\nu$, the allocated sum power increases as $\gamma$ decreases.  Intuitively, a constant rate constraint \eqref{eq:GammaProof_R} requires (\ref{eq:GammaProof_P}) to increase because the term $[\nu^*(\gamma)-\frac{1}{\beta_n}]^+$ increases faster than $[\log(\nu^*(\gamma)\beta_n)]^+$.  The following lemma shows this rigorously.

\begin{lem}
\label{lem:DecreasingInGamma}
The sum power $\sum_n p_n$ associated with allocation $[\gamma,\nu^*(\gamma)]$ is a decreasing function in $\gamma$.
\end{lem}
\begin{proof}
see appendix.
\end{proof}

Lemma~\ref{lem:DecreasingInGamma} enables us to find $\gamma^*$ again by the bisection method.  Assuming that \textbf{P1} is feasible which we will assume hereafter, there exist bounds $\gamma_u$ and $\gamma_l$ such that $P(\gamma_u,\nu^*(\gamma_u))\leq P\leq P(\gamma_l,\nu^*(\gamma_l))$.  Therefore, by starting the bisection method from these points we can find the pair $[\gamma^*,\nu^*]$ with guaranteed convergence.  The solution algorithm is shown in detail in Fig.~\ref{fig:BinarySearchAlgorithm}.  The inner loop (lines~4--14) correspond to finding $\nu^*(\gamma)$, whereas the outer loop finds $\gamma^*$.

\begin{figure}
\centering
\ifthenelse{\value{sgSingleColumn}=0}{%
\begin{minipage}{.65\linewidth}}{}
\restylealgo{ruled}\linesnumbered
\begin{algorithm}[H]
\SetLine
\textbf{Initialization.} Obtain bounds $\numin,\numax,\gamma_l,\gamma_u$\;
\Repeat{$0\leq P-\sum_n p_n\leq\epsilon_p$}{
$\hat{\gamma}\leftarrow(\gamma_u-\gamma_l)/2$\;
    \Repeat{$0\leq R-r(p_n,\rho_n)\leq\epsilon_R$}{
    $\hat{\nu}\leftarrow(\numax-\numin)/2$\;
    Find time allocation $\rho_n(\hat{\nu})$ using (\ref{eq:OptSolutionStructure-Rho})\;
    Find power allocation $p_n(\hat{\gamma},\hat{\nu})$ using (\ref{eq:OptSolutionStructure-P})\;
    Compute achievable rate $r(p_n,\rho_n)$ using (\ref{eq:AchievableRate})\;
    \eIf{$r(p_n,\rho_n)\geq R$}{$\numax\leftarrow\hat{\nu}$}{$\numin\leftarrow\hat{\nu}$}
    }
Find time allocation $\rho_n(\hat{\gamma},\hat{\nu})$ using (\ref{eq:OptSolutionStructure-Rho})\;
Find power allocation $p_n(\hat{\gamma},\hat{\nu})$ using (\ref{eq:OptSolutionStructure-P})\;
\eIf{$\sum_n p_n$}{$\gamma_l\leftarrow\hat{\gamma}$}{$\gamma_u\leftarrow\hat{\gamma}$}
}
\caption{Solution Algorithm}
\end{algorithm}
\ifthenelse{\value{sgSingleColumn}=0}{%
\end{minipage}}{}
\caption{Algorithm for finding the optimal Lagrange multipliers $\gamma$ and $\nu$ for problem (\ref{eq:Overlap-Objective})-(\ref{eq:Overlap-RegularityRho}).  The inner loop (lines 4--14) find $\nu^*(\gamma)$ which satisfies the rate constraint (\ref{eq:Overlap-RateConstr}).  The outer loop determines $\gamma^*$, which satisfies the power constraint (\ref{eq:Overlap-PowerConstr}).}
\label{fig:BinarySearchAlgorithm}
\end{figure}

\subsection{Properties of optimal allocations}
\label{sec:3c-StructuredResults}
Beyond simplifying solution algorithms, the structured solutions also enable us to make some qualitative statements about the optimal resource allocation.

\begin{figure*}
\centering
\subfloat[Ordering by IS channel quality]{\includegraphics[width=.48\linewidth]{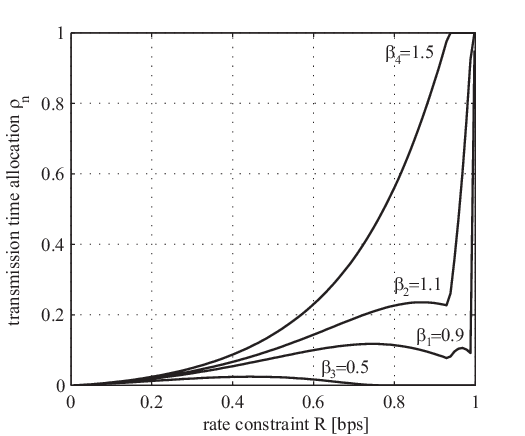}
\label{fig:3a-StructuredResultA}}
\subfloat[Ordering by sensing outcome]{\includegraphics[width=.48\linewidth]{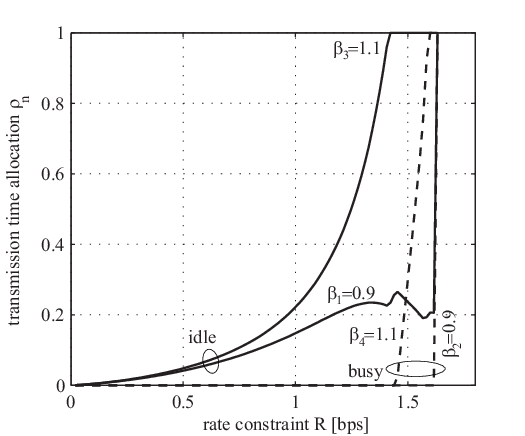}
\label{fig:3b-StructuredResultB}}
\caption{Structure of the optimal transmission time allocation for varying rate constraint.  The solution can be ordered with respect to IS channel coefficients or sensing outcomes. }
\label{fig:3-StructuredResults}
\end{figure*}

\subsubsection*{Ordering in terms of IS channel quality}
We first investigate the ordering with respect to the IS channel coefficients $\beta_n$.  This scenario is shown in Fig.~\ref{fig:3a-StructuredResultA} for $N=4$ sub-channels with coefficients $\bbeta = [.9, 1.1, .5, 1.5]^T$.  We observe that, for any rate constraint, more transmission time is allocated to the sub-channel with higher channel coefficient, \ie, $\beta_i\geq\beta_j\Rightarrow \rho_i\geq\rho_j$ provided all other parameters and the sensing results are identical.  On an intuitive level, this result captures the fact that in channels with high $\beta_i$ we can achieve the same rate in a shorter transmission duration using the same amount of power.  Mathematically, the result follows from the monotonicity of the optimal solution and $h_n(\nu)$.

\subsubsection*{Ordering in terms of sensing results}
Similar to the IS channel, the optimal transmission time allocation can be ordered with respect to the sensing results.  If channels have the same IS channel coefficient, $\beta_i=\beta_j$ but different sensing results then it is preferable to allocate more transmission time to the idle channel, \ie, $\rho_i\geq\rho_j$.  This is illustrated in Fig.~\ref{fig:3b-StructuredResultB} for $N=4$, $\bbeta=[.9, .9, 1.1, 1.1]^T$, and $\y=[0, 1, 0, 1]^T$.  It is also interesting to note that some transmission time is allocated to frames with busy sensing results even when idle frames are not yet used to the maximum extent.

\section{Optimal Average Resource Allocation}
\label{sec:4--AverageResourceAllocation}
Problem formulation \textbf{P1} required that rate and power constraint are met in every frame, even if sensing outcome or IS channel quality are disadvantageous.  In practical systems, satisfying rate constraints at the frame-level is usually unnecessary; it suffices to maintain average rate constraints across time.  This less stringent requirement can be used to further reduce interference by allocating less transmission time during frames with adversarial channel/interference conditions, while compensating for the rate decrease during frames with better conditions.  Ultimately, this leads to an improved resource allocation across both frequency (the sub-channels of the IS system) and time (consecutive frames of the IS system).

This section introduces such an average rate formulation by averaging across the temporal activity of the AH network and random IS channel coefficients.  Further, this section introduces two reference schemes that help to put the performance of the optimal resource allocation in perspective.

\subsection{Formulation and solution structure}
\label{sec:4a-FormulationSolutionStructure}
The average rate formulation requires associating probabilities with all possible sensing outcomes.  While there are a total of $N$ sub-channels available, the sensing outcomes for sub-channels that overlap with the same AH band will be identical.  Therefore, for $M$ sub-bands, there are a total of $2^M$ possible sensing outcomes.  Let the set of all possible sensing outcomes be represented by $\Y=\{0,1\}^M$ where $\y=[y_1,\ldots,y_M]^T\in\Y$ denotes the sensing outcome per sub-band.

Problem \textbf{P2} of optimally allocating power and transmission time then becomes
\begin{align}
\min\limits_{\substack{p_{n,\y}\\ \rho_{n,\y}}}\ & \sum_{\y\in\Y}\eta_\y\sum_n \phi_{n,y_{g(n)}}(\rho_{n,\y})\label{eq:AveSense-Objective} \\
\text{s.t.}\ & \sum_{\y\in\Y}\eta_\y\sum_n \rho_{n,\y}\log\left(1+\frac{p_{n,\y}\beta_n}{\rho_{n,\y}}\right) \geq R \label{eq:AveSense-RateConstr}\\
& \sum_{\y\in\Y}\eta_\y \sum_n p_{n,\y}  \leq P\label{eq:AveSense-PowerConstr}\\
& p_{n,\y}\geq 0,\quad \forall \y\in\Y, 1\leq n\leq N\label{eq:AveSense-RegularityP}\\
& 0\leq\rho_{n,\y}\leq 1,\quad \forall \y\in\Y, 1\leq n\leq N\label{eq:AveSense-RegularityRho},
\end{align}
where $\eta_{i,0}=\mu_i/(\lambda_i+\mu_i)$, $\eta_{i,1}=\lambda_i/(\lambda_i+\mu_i)$, and due to the independence of the AH sub-bands, $\eta_\y = \prod_{i=1}^M \eta_{i,y_i}$.  Note that this optimization problem has $2^M$ as many decision variables because power and transmission time allocation may be different for every possible sensing outcome.  The fact that the decision variables grow exponentially with $M$ is not of major concern, because $M$ (the number of parallel AH bands) is typically quite small (in the order of one to five).

Problem \textbf{P2} can be solved similar to problem \textbf{P1}.  In particular, by forming the Lagrangian, introducing Lagrange multipliers $\gamma$ and $\epsilon$, and taking the derivative with respect to the decision variables, we obtain a similar solution structure as in the frame-level problem.

\begin{figure*}[t!]
\centering
\subfloat[Frame length $T=1$]{\includegraphics{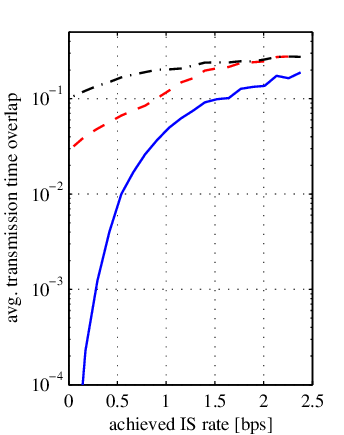}
\label{fig:4a-AverageResultA}}
\subfloat[Frame length $T=.1$]{\includegraphics{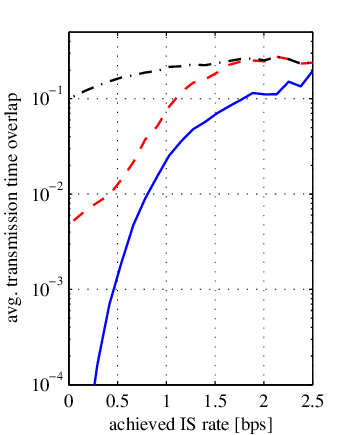}
\label{fig:4b-AverageResultB}}
\subfloat[Outage probability]{\includegraphics{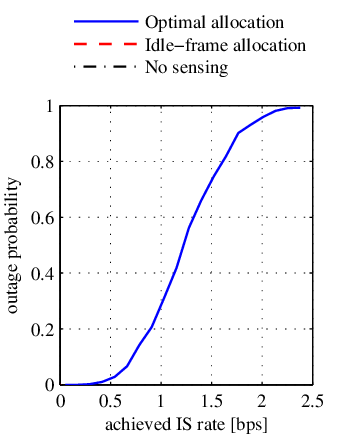}
\label{fig:4c-AverageResultC}}
\caption{Performance of optimal average rate resource allocation and comparison with suboptimal reference schemes. }
\label{fig:4-AverageResults}
\end{figure*}

\subsection{Reference schemes}
Having incorporated random sensing outcomes into our formulation, we introduce two reference schemes in order to put the performance of the optimal resource allocation in perspective.

\subsubsection*{No sensing case}
As a first benchmark, consider an approach that allocates power but does not perform any transmission time optimization.  This case corresponds to conventional resource management in IS systems, which simply allocates power to make the best use of the uplink channel.  Mathematically, this is formulated as minimizing $\sum_n p_n$ subject to the constraints (\ref{eq:Overlap-RateConstr})-(\ref{eq:Overlap-RegularityRho}).  We assume that for any sub-channel with $p_n>0$, the sub-channel is used for the entire frame duration.  Unused carriers for which $p_n=0$ are not allocated any transmission time.

\subsubsection*{Idle-frame allocation}
Another possible reference scheme performs spectrum sensing but allocates resources in a suboptimal way.  Specifically, consider allocating the entire frame by setting $\rho_n=1$ for all idle sub-channels while completely avoiding busy sub-channels by setting $\rho_n=0$ for all $n$ with $y_n=1$.  In the average rate formulation, this method can be formulated mathematically as minimizing $\sum_{\y\in\Y}\eta_\y\sum_n p_{n,\y}$
subject to (\ref{eq:AveSense-RateConstr})-(\ref{eq:AveSense-RegularityRho}) and the additional constraint that sub-channels with busy sensing outcome are never allocated.  Note that the above optimization problem may be infeasible even when \textbf{P2} is feasible because we are imposing the additional restriction of never transmitting during busy frames.  To ensure that the reference scheme is always feasible when \textbf{P2} is, we force allocation to busy channels if the resulting optimization would otherwise be infeasible.

\subsection{Allocation for random IS channels}
The previous section considered average rate constraints with respect to temporal AH activity but fixed IS channel $\bbeta$.  This section further extends the analysis to the case of random IS channel coefficients $\bbeta$.  The optimization problem \textbf{P3} is given by
\begin{align}
\min\limits_{\substack{\p(\y,\bbeta)\\ \brho(\y,\bbeta)}}\ & \int\limits_{\bbeta} \sum_{\y\in\Y}\eta_\y\sum_n \phi_{n,y_{g(n)}}(\rho_n(\y,\bbeta))dF(\bbeta)\label{eq:AveBeta-Objective} \\
\text{s.t.}\ & \int\limits_{\bbeta} \sum_{\y\in\Y}\eta_\y\sum_n \rho_n(\y,\bbeta)\log({\textstyle 1+\tfrac{p_n(\y,\bbeta)\beta_n}{\rho_n(\y,\bbeta)}})dF(\bbeta) \geq R \label{eq:AveBeta-RateConstr}\\
& \int\limits_{\bbeta} \sum_{\y\in\Y}\eta_\y \sum_n p_n(\y,\bbeta)dF(\bbeta)  \leq P\label{eq:AveBeta-PowerConstr}\\
& p_n(\y,\bbeta)\geq 0,\quad \forall \y\in\Y, 1\leq n\leq N\label{eq:AveBeta-RegularityP}\\
& 0\leq\rho_n(\y,\bbeta)\leq 1,\quad \forall \y\in\Y, 1\leq n\leq N,\label{eq:AveBeta-RegularityRho},
\end{align}
where the decision variables $\p(\y,\bbeta)$ and $\brho(\y,\bbeta)$ correspond to the power and transmission time allocation that is used for sensing outcome $\y$ and IS channel condition $\bbeta$ and $F(\bbeta)$ is the cumulative distribution function of $\bbeta$.
By again forming the Lagrangian and computing the derivative with respect to the decision variables, it is easy to show that the structured solutions (\ref{eq:OptimalityCondition-P}) and (\ref{eq:OptSolutionStructure-Rho})-(\ref{eq:OptSolutionStructure-RhoBusy}) again hold.  Therefore, it is again possible to express the allocation as a function of the Lagrange multipliers $[\gamma,\nu]$.  The rate constraint can then be evaluated by
\beq
\int\limits_{\bbeta}\sum_{\y\in\Y}\eta_{\y}\sum_n \rho_n(\y,\bbeta)[\log(\nu\beta_n)]^+dF(\bbeta)
\eeq
and the allocated sum power is given by
\beq
\int\limits_{\bbeta}\sum_{\y\in\Y}\eta_{\y}\sum_n \rho_n(\y,\bbeta)\left(\nu-\tfrac{1}{\beta_n}\right)^+dF(\bbeta).
\eeq
While the above integrals can only be evaluated numerically, it is possible to again find the optimal solution via the bisection method.

\subsection{Numerical results}
This section presents numerical performance results for the optimal average rate resource allocation and compares them to the reference schemes introduced in this section.  The results were obtained for $N=5$ sub-channels and a single AH sub-band $M=1$.  The prediction parameters were $\lambda=\mu=1\,$s$^{-1}$ and the IS channel coefficients were flat Rayleigh fading and statistically independent.  We further assume a block fading scenario in which the IS channel varies slowly compared to the frame duration.

The performance for fixed IS channel and random sensing results is shown in Fig.~\ref{fig:4a-AverageResultA} for $T=1\,$s and in Fig.~\ref{fig:4b-AverageResultB} for $T=.1\,$s.  The plot shows the average transmission overlap between IS and AH network versus the achieved IS rate (note that the achieved IS rate and not the rate constraint is plotted).  The performance results are averaged over 100 realizations of the IS channel.  Since the IS channel is not modeled statistically, it is inevitable that for some realizations of $\bbeta$ problem \textbf{P2} is infeasible.  The outage probability, which is identical for all three schemes, is therefore shown in Fig.~\ref{fig:4c-AverageResultC} to put the results in perspective.  Typical outage probabilities of approximately 10\% correspond to an IS rate of about 0.7$\,$bps.  At this rate, plots (a) and (b) show that a significant performance gain is achieved by performing sensing-based transmission time allocation.

The performance ordering reflects our expectations.  The idle-frame allocation scheme outperforms the no-sensing case but shows a quite significant performance gap with respect to the optimal allocation, especially for low IS rates.  Further, all curves show increasing interference as the IS rate increases.  This is expected, since high IS rates prevent the IS system from being able to accommodate the AH links.  The plots also show that idle-frame allocation and no-sensing scheme converge for high IS rates, because allocating only idle frames is almost always infeasible (and therefore busy frames typically need to be used as well).

By comparing Fig.~\ref{fig:4a-AverageResultA} ($T=1$) and Fig.~\ref{fig:4b-AverageResultB} ($T=.1$), we observe that while the performance of the optimal scheme does not change significantly, the idle-frame reference performs much better.  This is intuitive, because by reducing the frame length, it is easier to ``fill up'' the idle periods of the AH network.  The performance of the no-sensing scheme remains unaltered and is the same in both figures.

\begin{figure}[t]
\centering
\includegraphics[width=\linewidth]{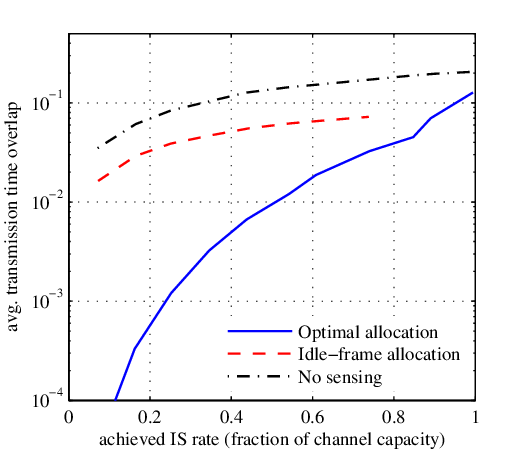}
\caption{Performance result for random IS channel coefficients.}
\label{fig:AveBeta}
\end{figure}

The performance for average IS channel coefficients and random AH behavior is shown in Fig.~\ref{fig:AveBeta} which compares the solution of \textbf{P3} with the same reference schemes.  We can observe that by exploiting the channel variability and allocating across frequency and time, we can further reduce interference.  Otherwise, the performance trends are similar to those of Fig.~\ref{fig:4-AverageResults}.  Note that idle-frame allocation does not achieve the same channel capacity as the optimal scheme because it only transmits in frames with an idle sensing result.

\section{Allocation for Multiple IS Users}
\label{sec:5--MultiTerminalAllocation} 
The previous section derived the optimal power and transmission time allocation assuming that an orthogonal set of sub-channels had already been assigned to each IS user.  This enabled us to consider each of the terminals individually and perform resource allocation based on local sensing results.

In practice, the IS base station needs to assign sub-channels to each of the IS users without knowing what the sensing outcomes will be.  We therefore consider the problem of optimal sub-channel allocation based on minimizing average interference metrics.  This leads to a similar formulation as compared to the average interference case in Sec.~\ref{sec:4--AverageResourceAllocation}.  Once a sub-channel allocation has been computed and fed back to the IS users, they can use the locally available sensing results to optimize their medium access.
The optimal sub-channel allocation is a combinatorial problem, which is computationally more challenging than problems \textbf{P1} through \textbf{P3}, which could be analyzed based on convex optimization.  While a general analysis of this problem goes beyond the scope of this paper we find the optimal solution for fairly small problem instances by exhaustive search.  A comparison with heuristic allocations suggests that efficient greedy sub-channel allocation algorithms developed for related problem setups can be adopted to this problem and yield a performance close to optimal.

\subsection{Optimal sub-channel allocation}
\begin{figure}
\centering
\includegraphics{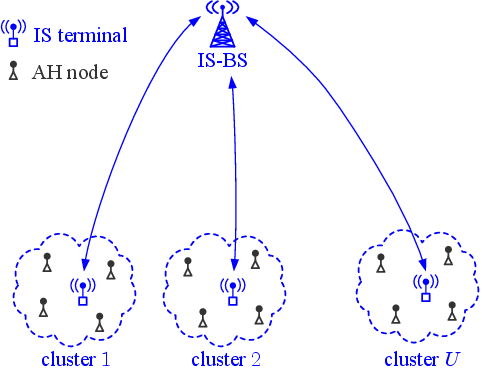}
\caption{System setup for the multi-terminal case.}
\label{fig:MultiUserSetup}
\end{figure}

The problem of optimal sub-channel allocation involves assigning orthogonal sets of sub-channels to each terminal, such that the overall interference is minimized; see Fig.~\ref{fig:MultiUserSetup}.  A mathematical formulation can be based on problem \textbf{P2}.  Specifically, define $f(\A)$ as the optimal value of \textbf{P2} where the summations over sub-channels are restricted to $n\in\A$,\ie, the sub-channels on which a specific IS user operates.
Define $f(\A):=\infty$ if \textbf{P2} is infeasible.

Assume that the IS base station is serving a total of $U$ users.  The problem of optimal sub-channel allocation is then formulated as
\begin{align}
\min\limits_{\{\A_u\}} & \sum_{u=1}^U f(\A_u)\\
\text{s.t.}\ & \A_1\cup\cdots\cup\A_U = \{1,\ldots,N\}\label{eq:Combinatorial1}\\
& \A_i\cap\A_j=\emptyset\quad i\neq j\label{eq:Combinatorial2},
\end{align}
where $\A_i$ denotes the set of sub-channels assigned to terminal $i$.  Due to \eqref{eq:Combinatorial1} and \eqref{eq:Combinatorial2} the sub-channel allocations are mutually exclusive and collectively exhaustive.  Note that the above problem does not require knowledge of the sensing outcome at the individual terminals.  The base station only requires knowledge of the CTMC parameters $\lambda$ and $\mu$, as well as knowledge of the IS channel coefficients $\bbeta$.

The above problem is difficult to solve due to its combinatorial nature and conventional sub-channel allocation methods are not easily extended to incorporate the additional dimension of allocating transmission time.  For small problem instances, however, the optimal allocation can be found by exhaustive search.

\subsection{Suboptimal algorithm}
The problem of optimal sub-channel allocation in multicarrier systems has been well-studied in the absence of transmission time allocation.  Standard methods typically minimize the total transmit power subject to rate constraints.  In our setup, this can be formulated mathematically as minimizing $\sum_{u=1}^U\sum_{n\in\A_u}p_n$ subject to rate and power constraints for the individual terminals.  While the resulting optimization problem is still combinatorial, efficient approximation techniques have been developed with close-to-optimal performance.

Allocating sub-channels in this manner, can be used as an effective heuristic. Since good channel quality results in lower average transmission time, we conjecture that conventional sub-channel allocation may be a good approximation to the optimal interference-aware sub-channel allocation.  Numerical results show that this is indeed the case in the scenarios we have examined.

\subsection{Numerical results}
\begin{figure}
\centering
\includegraphics[width=\linewidth]{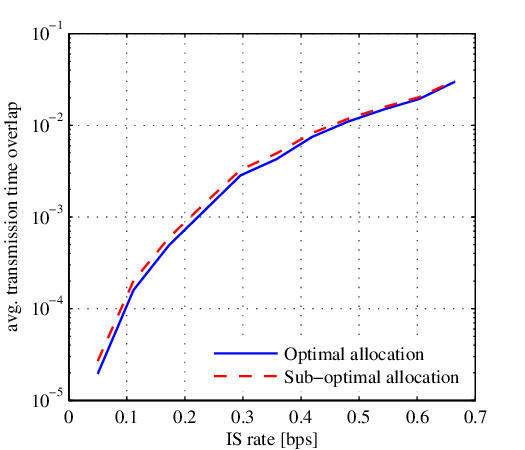}
\caption{Performance result for the multi-terminal case. }
\label{fig:MultiUser}
\end{figure}

Numerical results for the multi-terminal case are shown in Fig.~\ref{fig:MultiUser}.  The total average transmission time overlap (summed over all IS users) is plotted with respect to the rate constraint for each individual IS users (constraints are assumed to be identical).  The performance trends are the same as in the case of a single IS user.  For low rate requirements we can effectively mitigate interference by assigning resources judiciously.  On the other hand, as rate requirements become more stringent, there is less flexibility in accommodating the AH links.  The scenario plotted in Fig.~\ref{fig:MultiUser} corresponds to $U=3$ terminals, $N=5$ sub-channels, and flat Rayleigh fading IS channel coefficients.

The performance of the optimal and suboptimal sub-channel allocation schemes is very similar regardless of the rate constraint.  This suggests that minimizing the total transmission power is a reasonable approximation to the optimal sub-channel allocation.
In future work, we plan to corroborate this conjecture in more detail.

\section{Conclusion}
In conclusion, this paper has introduced a novel \emph{cognitive coexistence} framework, which enables infrastructure systems to coexist with local, ad-hoc or peer-to-peer communication links.  Based on sensing and predicting the interference patterns of these  ad-hoc links, the infrastructure system allocates power and transmission time judiciously such that overlaps with the ad-hoc links are minimized.

We analyzed the problem within the framework of convex optimization and derived closed-form solutions at the frame-level.  These results were extended to the average rate case to reduce interference further by allocation across both frequency and time.  Finally, we addressed the case of multiple IS users and  provided more insight on how sub-channel allocation can be performed.

\def\I{\mathbb{I}}

\appendices
\section{Proof of Lemma~1}
\label{app:ProofOfLemma1}
We denote the IS transmissions within the current frame by a finite set of closed and disjoint time intervals $\I_k=[a_k,b_k]$ where each $\I_k\subseteq [0,T]$ corresponds to a contiguous transmission of the IS user.  Clearly, this formulation incorporates possible pauses between IS transmissions.  We also require $\sum_k b_k-a_k=\rho T$, because a total of $\rho T$ transmission time needs to be allocated.

First, consider the case of an idle sensing result at the beginning of the frame, say at time $t=0$.  Then according to \eqref{eq:CostFunction-Idle} and \eqref{eq:CTMCTransitionMatrix}, the expected time overlap is given by
\ifthenelse{\value{sgSingleColumn}=0}{
\begin{multline}
\frac{1}{T}\sum_k\int_{a_k}^{b_k}\Pr(X(\xi)=1|X(0)=0)d\xi=
\frac{1}{T}\sum_k\int_{a_k}^{b_k}\frac{\lambda}{\lambda+\mu}\left(1-e^{-(\lambda+\mu)\xi}\right)d\xi.
\end{multline}
}{
\begin{multline}
\frac{1}{T}\sum_k\int_{a_k}^{b_k}\Pr(X(\xi)=1|X(0)=0)d\xi=\\
\frac{1}{T}\sum_k\int_{a_k}^{b_k}\frac{\lambda}{\lambda+\mu}\left(1-e^{-(\lambda+\mu)\xi}\right)d\xi.
\end{multline}}
Since the integrand is strictly increasing in $\rho$, the above expression is minimized by transmitting contiguously during the time interval $[0,\rho T]$.

In the case of a busy sensing result, an equivalent approach leads to a strictly decreasing integrand and therefore it is optimal to transmit during the time interval $[(1-\rho)T,T]$ in that case.

\section{Proof of Lemma~3}
In Sec.~\ref{sec:3b-IterativeSolutionAlgorithm} we have defined the sum power corresponding to the pair of Lagrange multipliers $[\gamma,\nu]$ as $P(\gamma,\nu)$.  Further, we showed that by keeping $\gamma$ fixed and varying $\nu$ it is possible to find a $\nu^*(\gamma)$ for which the rate constraint is satisfied with equality.  To simplify notation let us now define $P(\gamma)$ as the sum power associated with $[\gamma,\nu^*(\gamma)]$.

The proof that $P(\gamma)$ decreases with $\gamma$ proceeds by contradiction.  First, we note that $\gamma\rightarrow 0$ implies $P(\gamma)\rightarrow\infty$ due to the structure of the optimal solutions \eqref{eq:OptSolutionStructure-Rho}-\eqref{eq:OptSolutionStructure-RhoBusy}.  Assume now that $P(\gamma)$ is not monotonically decreasing.  Then, because $P(\gamma)$ is continuous, there exist two different values of $\gamma$, say $\gamma_1$ and $\gamma_2$, such that $P(\gamma_1)=P(\gamma_2)$.

Based on the KKT conditions stated in Sec.~\ref{sec:3a-OptimalityConditionsAndSolutionStructure}, it is easy to verify that both $\gamma_1$ and $\gamma_2$ correspond to optimal solutions of Problem~\textbf{P1} with rate constraint $R(\gamma_1,\nu^*(\gamma_1))$ and power constraint $P(\gamma_1,\nu^*(\gamma_1))$.  Further, from the structure of the optimal solutions it is clear that the transmission time allocations associated with $\gamma_1$ and $\gamma_2$ must be different, that is, $\brho(\gamma_1)\neq\brho(\gamma_2)$.  This is a contradiction, however, because Problem~\textbf{P1} has a strictly convex objective function and therefore at most one optimal solution. 

\bibliographystyle{IEEEtran}
\bibliography{IEEEabrv,refs,GlobeCom08}

\end{document}